\documentstyle[12pt]{article}
\def\abstract#1{{\centerline{\bg Abstract}} \vskip 3mm \par #1}
\def\cy{Calabi-Yau}
\def\cym{Calabi-Yau manifold}
\def\lg{Landau-Ginzburg}
\def\lgo{Landau-Ginzburg orbifold}

\def\tg{\tilde{\theta_{i}}^{g}}

\def\thd{\tilde{\theta_{i}}^{h^{\prime}}}

\def\inbar{\vrule height1.5ex width.4pt depth0pt} 
\def\ZZ{\relax{\sf Z\kern-.4em \sf Z}}  \def\IR{\relax{\rm I\kern-.18em R}}
\def\IN{\relax{\rm I\kern-.18em N}} \def\IP{\relax{\rm I\kern-.18em P}}
\def\IQ{\relax\,\hbox{$\inbar\kern-.3em{\rm Q}$}}
\def\la{\lambda}
\def\IC{\hbox{\,$\inbar\kern-.3em{\rm C}$}}

\def\ket#1{\left| #1\right\rangle}

\def\({\lbrack}
\def\){\rbrack}

\def\ketc#1{{\left| #1\right\rangle}_{\rm (c,c)}}
\def\keta#1{{\left| #1\right\rangle}_{\rm (a,c)}}

\begin{document}
\baselineskip=6mm
\begin{flushright}
{KOBE-TH-94-03} \\
{September 1994}
\end{flushright}
\vskip 1.5cm
\centerline{\LARGE {\bf The Monomial-Divisor Mirror Map for}}
\vskip 1cm
\centerline{\LARGE {\bf Landau-Ginzburg Orbifolds}}
\vskip 3.5cm
\centerline{\large Hitoshi \ Sato}
\vskip 1cm
\centerline{\it Graduate School of Science and Technology, Kobe University}
\centerline{\it Rokkodai, Nada, Kobe 657, Japan}
\centerline{email address : UTOSA@JPNYITP.BITNET}
\vskip 2.5cm
\centerline{\large {\bf ABSTRACT} }

\vskip 0.5cm

We present the new explicit geometrical knowledge
of the \lg\ orbifolds,
when a typical type of superpotential is considered.
Relying on toric geometry,
we show the one-to-one correspondence
between some of the $(a,c)$ states with $U(1)$ charges
$(-1,1)$ and the $(1,1)$ forms coming from
blowing-up processes.
Consequently, we find
the monomial-divisor mirror map for \lg\ orbifolds.
The possibility of the application of
the models of other types
is briefly discussed.

\thispagestyle{empty}

\clearpage
\pagenumbering{arabic}



$N=2$ superconformal field theory has attracted
the attention in the context of
string compactification \cite{g1}.
Due to its (anti-)chiral ring structure \cite{lvw,dixon} ,
the theory with $c=9$ has a \cy\ interpretation,
i.e. the $(p,q)$ forms on a \cym\
can be identified with
$(3-p,q)$ states of the $(c,c)$ ring
or $(-p,q)$ states of the $(a,c)$ ring,
where $c \ (a)$ stands for (anti-)chiral
and the states are labeled by the $U(1)$ charges.
These $(c,c)$ and $(a,c)$ rings can be described
in terms of the \lg\ models.

Recently, the mirror symmetry of \cym s has been
actively studied \cite{cop,hkty1},
since some Yukawa couplings can be determined exactly,
assuming that this symmetry is correct.
Although this symmetry was first suggested in
the $N=2$ superconformal field theory context \cite{lvw,dixon} ,
recent analysis is purely geometrical.



Toric geometry gives us the method to examine some of the moduli
of \cy\ manifolds.
These are the $(1,1)$ forms coming from the blowing-up process
and the $(2,1)$ forms which are mirror partners of them.
These $(2,1)$ forms can be represented by monomials
in a defining equation of a \cym .
Aspinwall et.al.\cite{agm3} found that
these $(2,1)$ and $(1,1)$ forms get interchanged
under the mirror map.
Hence this mirror map is called
``{\it the monomial-divisor mirror map}''.


In this paper, we try to find the corresponding mirror map
in the \lg\ context.
To do this, we first identify the $(-1,1)$ states
with the $(1,1)$ forms coming from the blowing-up process.
Once this identification is made,
we can study the geometry of compactified space
more deeply in terms of \lg\ model.

In this paper, we will restrict our attention to
the superpotential of a form
$W{(X_{i})} =
X_{1}^{a_{1}}+X_{2}^{a_{2}}+X_{3}^{a_{3}}
+X_{4}^{a_{4}}+X_{5}^{a_{5}},$
which corresponds to the Fermat type hypersurface
in $WCP^{4}$.
The \lg\ orbifolds are obtained by
quotienting with an Abelian symmetry group $G$ of
$W{(X_{i})}$,
whose element $g$ acts as an $N \times N$ diagonal matrix,
$g: X_{i} \rightarrow e^{2 \pi i {\tg}}X_{i}$,
where $0 \leq \tg < 1$.
Of course the $U(1)$ twist \
$j: X_{i} \rightarrow e^{2 \pi i {q_{i}}}X_{i}$ \
generates the symmetry group of
$W{(X_{i})}$,
where $q_{i} = {n_{i} \over d}$,\ \
 $W(\la^{n_{i}} X_{i}) = \la^{d}W(X_{i})$ \ and
$\la \in \IC^{\ast}$.

Using the results of Intriligator and Vafa \cite{iv},
we can construct the $(c,c)$ and $(a,c)$ rings.
Also we could have the left and right $U(1)$ charges of
the ground state \
$\keta h$ \
in the $h$-twisted sector
of the $(a,c)$ ring.
In terms of spectral flow, \ $\keta h$ is mapped to the (c,c)
state $\ketc {h^{\prime}}$ with $h^{\prime} = hj^{-1}$.
Then the charges of the (a,c) ground state of $h$-twisted sector
$ \keta {h} $
are obtained to be
\begin{equation}
\label{uac}
\begin{array}{cc}
\left(\begin{array}{c}
J_{0} \\
\bar{J_{0}}
\end{array} \right) &
\end{array}
\keta {h}
=
\begin{array}{cc}
\left(\begin{array}{c}
{ - \sum_{\thd>0}{(1-q_{i}-\thd)}}
+ \sum_{\thd=0} {(2q_{i}-1)}
 \\
{ \sum_{\thd>0}{(1-q_{i}-\thd)}}
\end{array} \right) &
\keta {h}.
\end{array}
\end{equation}

Using this result, we see that the $(-1,1)$ states written in the from
$\keta {j^{l}}$ can always arise from the twisted sector
with $I^{\prime} = 0$
, where $I^{\prime}$ is the number of the invariant fields $X_{i}$
under the $h^{\prime}$ action.
{}From the results of ref.\cite{sa1}, we see that the $(2,1)$ states
corresponding to the $(-1,1)$ states
can come from the $h^{\prime}$ twisted sector
with $I^{\prime} = 0$ or $I^{\prime} = 2$.
So the $(-1,1)$ states can arise from the twisted sectors
with $I^{\prime} = 2$ only if ${\sum_{\thd=0}{(2q_{i}-1})} = 0$.
This condition implies that the \lg\ superpotential
contains two trivial fields.
So as long as we consider the \lg\ models
with no or one trivial field,
the $(-1,1)$ states which can be represented by
$\keta {j^{l}}$ may exist only in the twisted sector
with $I^{\prime} = 0$.

Let us turn our attention to geometry.
\cym s are represented by hypersurfaces in $WCP$.
In general, due to the $WCP$ identification\
$z_{i} \sim \la^{n_{i}}z_{i}$,\ \ $\la \in \IC^{\ast}$, \
we have some fixed sets on a hypersurface.
When we consider \cy\ $3$-folds,  possible fixed sets are
fixed points and fixed curves.
To obtain a smooth \cym\, we have to blow up these singularities.
Hodge numbers $h^{1,1}$ and $h^{2,1}$
change through the blowing-up processes.
Especially $h^{1,1}$ increases since new $(1,1)$ forms arise
from exceptional divisors,
which come from the resolution of
singularities.

Those \cy\ resolutions can be described in terms of toric geometry
\cite{ro3,vb,hkty1}.
Toric geometry describes the structure of a certain class of
geometrical spaces in terms of simple combinatorial data.
When a space admits a description in terms of toric geometry,
many basic and essential characteristics of the space -
such as its divisor classes and other aspects of its cohomology -
are neatly coded and easily deciphered from the analysis of
corresponding lattices.
In toric geometry, we are able to deal with some of the exceptional
divisors, which we call toric divisors.


First we consider a fixed curve in $WCP^{4}$.
We will briefly summarize the description of toric divisors
in terms of toric data following ref. \cite{ro3},
and explain the idea of the identification
in this case.

Let $G^{\prime}$ be a finite group generated by $g^{\prime}$
which acts on $z_{i}$, homogeneous coordinates of $WCP^{4}$, as

\begin{equation}
\label{gact1}
g^{\prime} \ : \ [z_{1}, z_{2}, z_{3},z_{4},z_{5} ] \rightarrow
[ e^{2 \pi i x_{1}}{z_{1}},e^{2 \pi i x_{2}}{z_{2}},
z_{3}, z_{4}, z_{5} ].
\end{equation}
The curve in $WCP^{4}$ fixed under the $g^{\prime}$ action
can be written in the form

\begin{equation}
z_{3}^{a_{3}}+z_{4}^{a_{4}}+z_{5}^{a_{5}} = 0,
\ \ z_{1} = z_{2} = 0.
\end{equation}

In this case let us define

\begin{equation}
\label{n1}
{\bf n} = \left\{ \pmatrix{
x_{1} \cr
x_{2} \cr
} \in \IR^{2} \mid {\rm dia} \left[
e^{2 \pi i x_{1}},
e^{2 \pi i x_{2}},
\right] \in G^{\prime}
\right\},
\end{equation}

\par

\begin{equation}
\label{tri1}
{\bf \triangle} = \left\{ \pmatrix{
x_{1} \cr
x_{2} \cr
} \in \IR^{2} \mid \sum_{i=1}^{2}{x_{i}} = 1, x_{i} \ge 0\
{\rm for\ all}\ i
\right\},
\end{equation}

\par

\begin{equation}
\Gamma = {\bf n} \cap {\bf \triangle}.
\end{equation}
$\Gamma$ is a finite subset of the lattice ${\bf n}$, and contains the
standard base ${\left\{ e^{i} \right\}}_{i=1}^{2}$ of $\IR^{2}$.
These are called toric data.

It is known that
\begin{eqnarray}
\left\{ D_{\gamma} \mid \gamma \in \Gamma -
{\left\{ e^{i} \right\}}_{i=1}^{2} \right\} \nonumber \\
& = & \left\{ {\rm toric \ divisors \ coming \ from \ resolution}
\right\}.
\end{eqnarray}

In the following
we associate a point in the lattice
$\Gamma$, i.e. a toric divisor, with a $(-1,1)$ state
which can be written in the form $\keta {j^{l}}$ with $I$ = 3,
 if $g^{\prime}$
can be written in the form $j^{l}$
, where $I$ is the number of the invariant fields $X_{i}$
under the $j^{l}$ action.

We define the phase symmetries $\rho_{i}$ which act on $X_{i}$ as
\begin{equation}
\rho_{i}X_{i} = e^{2 \pi i q_{i} }X_{i},
\end{equation}
with trivial action for other fields.
The operator $ \rho_{i} $
can be represented by a diagonal matrix
whose diagonal matrix elements are 1 except for
$ {(\rho_{i})}_{i,i}
= e^{2 \pi i q_{i} }$.

It is obvious that
\begin{equation}
j = \rho_{1} \cdots \rho_{5}.
\end{equation}
In the $j^{l}$-twisted sector, if a field $X_{i}$ is invariant
then
\begin{equation}
\rho_{i}^{l} = \rho_{i}^{l_{i}} = {\rm identity},
\end{equation}
where $l_{i} \equiv l \ {\rm mod} \ a_{i}$ and we have
\begin{equation}
j^{l} = \prod_{{l{q_{i}}} \notin \ZZ} {\rho_{i}^{l_{i}}}.
\end{equation}

If the number of $i$'s which satisfy $lq_{i} \notin \ZZ$
is 2, i.e. $I = 3$,
$j^{l}$ acts on $X_{i}$ as
\begin{equation}
\label{jact1}
j^{l} \ : \
[
X_{1}, X_{2}, X_{3},X_{4},X_{5} ] \sim
[ e^{2 \pi i l_{1} q_{1}}{X_{1}},
e^{2 \pi i l_{2} q_{2}}{X_{2}},
X_{3}, X_{4}, X_{5} ].
\end{equation}
with an appropriate renumbering for $q_{i}$, if necessary.
Then $j^{l}$ can be equivalent to $g^{\prime}$
through the identification
\begin{equation}
\label{itd1}
[e^{2 \pi i l_{1}q_{1}},e^{2 \pi i l_{2}q_{2}} ] \sim
[e^{2 \pi i x_{1}},e^{2 \pi i x_{2}} ].
\end{equation}
The condition for $\triangle$ is automatically satisfied
since $\det {j^{l}} = 1$.
Thus we can associate a (a,c) state
$\keta {\prod_{{lq_{i}} \notin \ZZ}{\rho_{i}^{{l_{i}}}}}$
with $I$=3 with
a point in $\Gamma$,
i.e. a toric divisor.

Further we can calculate the $U(1)$ charges of this state
using eq.(\ref{uac})
and the result is
\begin{equation}
{( - \sum_{i=1}^2{l_{i}q_{i}}, \sum_{i=1}^2{l_{i}q_{i}})} \sim
{( - \sum_{i=1}^2{x_{i}}, \sum_{i=1}^2{x_{i}})},
\end{equation}
through the identification (\ref{itd1}).
Thus we find that
the $U(1)$ charges of the state
$\keta {\prod_{{lq_{i}} \notin \ZZ}{\rho_{i}^{{l_{i}}}}}$
with $I$=3 is $(-1,1)$,
since the condition for $\triangle$ holds.

Now we can find the one-to-one correspondence
between $(-1,1)$ states and $(1,1)$ forms.
It is believed that $(-1,1)$ states correspond to
$(1,1)$ forms
and we know a new $(1,1)$ form comes from a toric divisor
during blowing up processes.
Therefore,
a charge $(-1,1)$ state
$\keta {\prod_{{lq_{i}} \notin \ZZ}{\rho_{i}^{{l_{i}}}}}$
with $I$=3
corresponds to a $(1,1)$ form coming from a toric divisor
through the identification (\ref{itd1}).

Also the $(2,1)$ states can be identified with
the $(1,1)$ forms.
By spectral flow, the ground state $\keta h$ is mapped to
the ground state $\ketc {h^{\prime}}$ with
$h^{\prime} = hj^{-1}$.
So we conclude that the charge $(-1,1)$ state $\keta h$
flows to the charge $(2,1)$ state $\ketc {h^{\prime}}$
in $h^{\prime}$-twisted sector with $I^{\prime} = 0$.
Thus a $(1,1)$ form coming from a toric divisor
corresponds to a charge (2,1) state $\ketc {h^{\prime}}$
in $h^{\prime}$-twisted sector with $I^{\prime} = 0$.

It is easy to see that the state $\ketc {j^{-2}}$
always satisfies this condition.
But the corresponding $(-1,1)$ state $\keta {j^{-1}}$
has $I = 0$.
We should associate this state  with the origin of
 the lattice $\Gamma$,
which corresponds to the $(1,1)$ form coming from the
embedded space $WCP^{4}$.

It is instructive to explain the above story by
a simple example.
Consider the \lg\ model with the superpotential
\begin{equation}
\label{w21}
W_{1} =
X_{1}^{8}+X_{2}^{8}+X_{3}^{4}+X_{4}^{4}+X_{5}^{4},
\end{equation}
with $U(1)$ charges of $X_{i}$ being
\begin{equation}
(
\frac{1}{8},
\frac{1}{8},
\frac{1}{4},
\frac{1}{4},
\frac{1}{4}
).
\end{equation}
The orbifold model $W_{1} / j$ has a corresponding $\ZZ_2$ fixed curve
which can be written
\begin{equation}
z_{3}^{4}+z_{4}^{4}+z_{5}^{4} = 0 \ \ \
{\rm in } \  WCP_{(1,1,2,2,2)}^{4}.
\end{equation}
After the blowing up only one (1,1) form comes from
a toric divisor.

Using the above discussions we can easily find the state
corresponding to this (1,1) form.
It is easy to see that in this case the operator
which acts like $g^{\prime}$ in (\ref{gact1}) is $j^{4}$.
So the twisted ground state $\keta {j^{4}}$ is associated with
the (1,1) form coming from toric divisor.
By flowing to the (c,c) ring, we find that
the state $\ketc {j^{3}}$ has the same correspondence.
It is worth noting that in this model there exists
the state $\keta {j^{-1}}$ (or $\ketc {j^{-2}}$)
which corresponds to the $(1,1)$ form coming from
$WCP^{4}$.

Next we consider fixed points in $WCP^{4}$.
The description of toric divisors
in terms of toric data is as follows \cite{ro3}.

Let $G^{\prime}$ be a finite group generated by $g^{\prime}$
which acts on $z_{i}$ as

\begin{equation}
\label{gact2}
g^{\prime} \ : \ [z_{1}, z_{2}, z_{3},z_{4},z_{5} ] \rightarrow
[ e^{2 \pi i x_{1}}{z_{1}},e^{2 \pi i x_{2}}{z_{2}},
e^{2 \pi i x_{3}}{z_{3}},
z_{4}, z_{5} ].
\end{equation}
The points in $WCP^{4}$ fixed under the $g^{\prime}$ action
can be written in the following form

\begin{equation}
z_{4}^{a_{4}}+z_{5}^{a_{5}} = 0,
\ \ z_{1} = z_{2} = z_{3} = 0.
\end{equation}

In this case the toric data are

\begin{equation}
{\bf n} = \left\{ \pmatrix{
x_{1} \cr
x_{2} \cr
x_{3} \cr
} \in \IR^{3} \mid {\rm dia} \left[
e^{2 \pi i x_{1}},
e^{2 \pi i x_{2}},
e^{2 \pi i x_{3}}
\right] \in G^{\prime}
\right\},
\end{equation}

\par

\begin{equation}
{\bf \triangle} = \left\{ \pmatrix{
x_{1} \cr
x_{2} \cr
x_{3} \cr
} \in \IR^{3} \mid \sum_{i=1}^{3}{x_{i}} = 1, x_{i} \ge 0\
{\rm for\ all}\ i
\right\},
\end{equation}

\par

\begin{equation}
\Gamma = {\bf n} \cap {\bf \triangle}.
\end{equation}
$\Gamma$ is a finite subset of the lattice ${\bf n}$, and contains the
standard base ${\left\{ e^{i} \right\}}_{i=1}^{3}$ of $\IR^{3}$.

In this case two classes of fixed points are possible.
One consists of the isolated fixed points and the other
consists of the fixed points on fixed curves.
In each type a toric divisor coming from the resolution
of fixed point is associated with a point in the lattice

\begin{equation}
\Gamma_{\rm in} = {\bf n} \cap {\rm interior} \ ({\bf \triangle}),
\end{equation}
which is a sublattice of $\Gamma$.
In general, the curve is fixed by the subgroup of the group
which fixes some points on that curve.
This subgroup can be reduced to the group $G^{\prime}$
in eq.(\ref{n1}), after appropriate
renumbering if necessary.
So the toric divisors coming from the resolution of
fixed curves on which fixed points sit
can be associated with the points in the sublattice of $\Gamma$.

For the toric divisors coming from the resolution of
fixed curves we can identify a $(1,1)$ form with a $(-1,1)$ state
$\keta {\prod_{{lq_{i}} \notin \ZZ}{\rho_{i}^{{l_{i}}}}}$
with $I$=3
in the same way.
Using a similar discussion we can identify
a $(1,1)$ form coming from the resolution of fixed point
with a $(-1,1)$ state
$\keta {\prod_{{lq_{i}} \notin \ZZ}{\rho_{i}^{{l_{i}}}}}$
with $I$=2.

Let us now turn to the mirror map.
It is known that the mirror of \lgo\ $W / j$ is obtained
to be $W / G_{m}$, where $G_{m}$ is the maximal
phase symmetry group of $W$ with determinant 1 \cite{gp}.
Although the mirror theory $W / G_{m}$ has the same potential
$W$ as the original theory $W / j$, we will denote it
by $\overline{W}$, which consists of the fields $\overline{X_{i}}$
(of course $\overline{X_{i}} = X_{i}$ in this case).
This is to make clear which theory we are considering.

Unfortunately, we cannot fully establish mirror pairings
of the states, but we can discuss the mirror partners of
a special type of states.
They are the states which can be written in the form
$\keta {\prod_{i}{{\rho}_{i}^{-l_{i}}}}$,
where $l_{i}$ are defined mod $a_{i}$ as before.

By using eq.(\ref{uac}), it can be shown that
the left $U(1)$ charge of the state
$\keta {\rho_{i}^{-1}}$ is $-q_{i}$ and right charge is $q_{i}$.
This fact suggests that the mirror image of the twisted
ground state $\keta {\rho_{i}^{-1}}$ is
${\overline{X_{i}}}\ket0$.
So we would like to conjecture
\begin{equation}
{\keta {{{\rho_{i}^{-{l_{i}}}}}} }
\buildrel {\rm mirror \  pair} \over \longleftrightarrow
{{{{\overline{X_{i}}}^{{l_{i}}}}}\ket0}.
\end{equation}
If we consider the more general twisted ground state
${\keta {\prod_{i}{{\rho_{i}^{-{l_{i}}}}}} }$,
we can write the above mirror pairing as
\begin{equation}
{\keta {\prod_{i}{{\rho_{i}^{-{l_{i}}}}}} }
\buildrel {\rm mirror \  pair} \over \longleftrightarrow
{{\prod_{i}{{\overline{X_{i}}}^{{l_{i}}}}}\ket0}.
\end{equation}

In terms of this mirror pairing we can find the mirror partner
of the $(-1,1)$ state which is discussed above.
Since this state can be represented by
$\keta {\prod_{{lq_{i}} \notin \ZZ}{\rho_{i}^{-{l_{i}}}}}$
we see that the mirror partner of this state is
${{\prod_{{lq_{i}} \notin \ZZ}{\overline{{X_{i}}}^{{l_{i}}}}}}\ket0$.

We should call this pairing
the monomial-divisor mirror map for \lgo\,
because this state
${{\prod_{{lq_{i}} \notin \ZZ}{\overline{{X_{i}}}^{{l_{i}}}}}}\ket0$
must correspond to the monomial
$\prod_{{lq_{i}} \notin \ZZ}{X_{i}^{{l_{i}}}}$
which survives the orbifoldization by $G_{m}$,
where we have omitted the bar over $X_{i}$.
The monomial-divisor mirror map for \cy\ mirror pair
is studied in \cite{hkty1,agm2},
and we have checked that our results exactly correspond to
the results obtained therein .

For example, we take the superpotential (\ref{w21}) again
(this model is considered in \cite{hkty1}).
In this example,
the twisted ground state $\keta {j^{4}}$ is associated with
the (1,1) form coming from resolution.
Since $j^{8} = 1$ we have
\begin{equation}
\keta {j^{4}}
\sim
\keta {j^{-4}}.
\end{equation}
Through the fact
\begin{equation}
{j^{-4}} = {\rho_{1}^{-4}}{\rho_{2}^{-4}},
\end{equation}
we find the mirror pairing

\begin{equation}
{\keta  {{\rho_{1}^{-4}}{\rho_{2}^{-4}} }
\buildrel {\rm mirror \  pair} \over \longleftrightarrow
{\overline{X_{1}}}^{4}
{\overline{X_{2}}}^{4}\ket0}.
\end{equation}
So we conclude that the monomial which survives after
the orbifoldization
by the $G_{m}$ action is
${{X_{1}}}^{4}
{{X_{2}}}^{4}$
, where we have omitted the bar over $X_{i}$.

We can find the mirror partner of the state
$\keta {j^{-1}}$
which corresponds to the $(1,1)$ form coming from $WCP^{4}$.
The mirror image is ${\overline{X_{1}}}
{\overline{X_{2}}}
{\overline{X_{3}}}
{\overline{X_{4}}}
{\overline{X_{5}}}
\ket0$.
This corresponds to the monomial
$X_{1}
X_{2}
X_{3}
X_{4}
X_{5}$
which is evidently invariant under the $G_{m}$ action.
We summarize these results in Table \ref{mdl1}
together with those for $(c,c)$ states.

\begin{table}[htbp]
\[ \begin{tabular}{||c|c|c||} \hline
(c,c) state & (a,c) state & mirror partner \\ \hline \hline
$\ketc {j^{-2}}$ & $\keta {j^{-1}}$ &
${{X_{1}}^{}}
{{X_{2}}^{}}
{{X_{3}}^{}}
{{X_{4}}^{}}
{{X_{5}}^{}}
\ket0 $ \\ \hline
$\ketc {j^{-4}}$ & $\keta {j^{-3}}$ &
${{X_{1}}^{4}}
{{X_{2}}^{4}}
\ket0 $ \\ \hline
\end{tabular} \]
\caption{The monomial-divisor mirror map for
\lg\ orbifolds of $W_{1}$}
\label{mdl1}
\end{table}

As a more complicated example we take the following
\lg\ superpotential
\begin{equation}
\label{w63}
W_{2} =
X_{1}^{3}+X_{2}^{3}+X_{3}^{6}+X_{4}^{9}+X_{5}^{18},
\end{equation}
with $U(1)$ charges
\begin{equation}
(
\frac{1}{3},
\frac{1}{3},
\frac{1}{6},
\frac{1}{9},
\frac{1}{18}
).
\end{equation}
This model is considered in \cite{agm2,agm1}.
The orbifold model $W_{2} / j$ has one corresponding
$\ZZ_2$ fixed curve,
one corresponding $\ZZ_3$ fixed curve
and  corresponding $\ZZ_6$ fixed points on the
intersections of these curves.
They can be written as
\begin{equation}
{\rm \ZZ_{2} \ fixed \ curve} \
z_{1}^{3}+z_{2}^{3}+z_{4}^{9} = 0
\end{equation}
\begin{equation}
{\rm \ZZ_{3} \ fixed \ curve} \
z_{1}^{3}+z_{2}^{3}+z_{3}^{6} = 0
\end{equation}
\begin{equation}
{\rm \ZZ_{6} \ fixed \ points} \
z_{1}^{3}+z_{2}^{3} = 0 \ \quad
{\rm in } \ WCP_{(6,6,3,2,1)}^{4}.
\end{equation}

It is easy to find the states which correspond to the
(1,1) forms coming from toric divisors
and their mirror partners.
The results are displayed in Table \ref{mdl2},
 where we have omitted the bar over $X_{i}$.

\begin{table}[htbp]
\[ \begin{tabular}{||c|c|c||} \hline
(c,c) state & (a,c) state & mirror partner \\ \hline \hline
$\ketc {j^{-2}}$ & $\keta {j^{-1}}$ &
${{X_{1}}^{}}
{{X_{2}}^{}}
{{X_{3}}^{}}
{{X_{4}}^{}}
{{X_{5}}^{}}
\ket0 $ \\ \hline
$\ketc {j^{-4}}$ & $\keta {j^{-3}}$ &
${{X_{3}}^{3}}
{{X_{4}}^{3}}
{{X_{5}}^{3}}
\ket0 $ \\ \hline
$\ketc {j^{-7}}$ & $\keta {j^{-6}}$ &
${{X_{4}}^{6}}
{{X_{5}}^{6}}
\ket0 $ \\ \hline
$\ketc {j^{-10}}$ & $\keta {j^{-9}}$ &
${{X_{3}}^{3}}
{{X_{5}}^{9}}
\ket0 $ \\ \hline
$\ketc {j^{-13}}$ & $\keta {j^{-12}}$ &
${{X_{4}}^{3}}
{{X_{5}}^{12}}
\ket0 $ \\ \hline
\end{tabular} \]
\caption{The monomial-divisor mirror map for
\lg\ orbifolds of $W_{2}$}
\label{mdl2}
\end{table}
This result agrees with the one obtained in \cite{agm2} .
Note that we do not need any geometrical informations
such as the number of fixed sets or the relations among them.

The corresponding \cym\ has five $(1,1)$ forms
discussed above
and two $(1,1)$ forms whose mirror partners
cannot be described by the monomials.
The $(-1,1)$ states corresponding to these two $(1,1)$ forms
are represented by
$X_{1} {\keta {j^{-2}}}$ \ \
and $X_{2} {\keta {j^{-2}}}$.
But we do not know the one-to-one correspondence
and their mirror partners.


Let us discuss the meaning of our results
a little bit more.
We used the toric data to identify $(a,c)$ states with $(1,1)$
forms.
This is a remarkable fact.
At this moment, it is unclear
why \lg\ models have \cy\ interpretations.
However, our method could partially answer
to this problem.
The toric data are essential and they have two different
interpretations,
i.e. cohomologies on a \cym\ and $(a,c)$ states in a \lg\ orbifolds.

Also, our method could answer to another important problem,
i.e. why strings do not feel singularities.
Our analyses show that the states coming from twisted sectors
correspond to the forms coming from
blowing-up processes.
Since the modular invariance of the Witten index requires
these twisted sectors,
we obtain the index as an Euler number of a smooth \cym .

The superpotential considered in this paper
corresponds to the Gepner model of A-type \cite{g1}.
So our analysis will give the insight into
the understanding of the exact mirror map
at the level of the conformal field theory.

Some problems still remain.
In this paper we restrict our attention to the
Fermat-type potential with five fields.
Of course the potentials of other types are possible
for string compactification.
For those potentials, there could be new singularities
whose resolutions cannot be described
in terms of toric geometry.
The $(-1,1)$ states, which correspond to the $(1,1)$ forms coming
from resolutions of these singularities of new type,
should not be written in the form
$\keta {j^{l}}$.
However, in general there can be $(-1,1)$ states
written in the form
$\prod_{{lq_{i}} \in \ZZ} {{X_{i}}^{l_{i}}} {\keta {j^{l}}}$
with at least one $l_{i} > 0$.
{}From the results of ref.\cite{sa1},
we see that corresponding $(2,1)$ states can
arise from the $j^{l-1}$-twisted sector with $I^{\prime} = 2$.
In ref.\cite{kreu}, non-Fermat-type potentials and
their mirror maps are considered.
But we can not fully establish the one-to-one correspondence
between the $(-1,1)$ states and the $(1,1)$ forms.

However, there are non-Fermat-type potentials with
only the singularities which can be treated through toric geometry.
For example, there is the hypersurface embedded in
$WCP_{(1,2,2,2,3)}^{4}$ ,
which gets two toric divisors after blowing up \cite{beka}.
For this model, our identification of the $(-1,1)$ states holds
and we have checked that the monomial-divisor mirror map for \lgo\
is indeed realized.

Although we do not know an exact method
for the calculation of the Yukawa couplings
in the framework of \lg\ models,
our analysis will be useful to study the
moduli dependence of the Yukawa couplings.

\vspace{1cm}

{\it Acknowledgements} :
The author would like to thank M. Ida and C.S. Lim
for helpful discussions and
for careful reading of this manuscript.


\end{document}